\documentclass[preprint]{aastex}
\usepackage{natbib}
\usepackage{placeins}
\usepackage{figs}

\newcommand{\tMASS}{${\it 2MASS }$}
\newcommand{\IRAC}{${\it IRAC }$}
\newcommand{\LGGS}{${\it LGGS }$}
\newcommand{\WISE}{${\it WISE }$}

\begin{document}
\title{Luminous and Variable Stars in M31 and M33. III. The  Yellow and Red Supergiants and Post-Red Supergiant Evolution}
\author{Michael S. Gordon, Roberta M. Humphreys, Terry J. Jones}
\affil{Minnesota Institute for Astrophysics}
\affil{116 Church St SE, University of Minnesota, Minneapolis, MN 55455, USA}
\email{gordon@astro.umn.edu, roberta@umn.edu, tjj@astro.umn.edu}
\altaffiltext{}{Based on observations  obtained with the Large Binocular Telescope (LBT), an international collaboration among institutions in the United  States, Italy and Germany. LBT Corporation partners are: The University of Arizona on behalf of the Arizona university system; Istituto Nazionale di Astrofisica, Italy; LBT Beteiligungsgesellschaft, Germany, representing the  Max-Planck Society, the Astrophysical Institute Potsdam, and Heidelberg  University; The Ohio State University, and The Research Corporation, on behalf of The University of Notre Dame, University of Minnesota and University of Virginia.}

\begin{abstract}
  Recent supernova and transient surveys have revealed an increasing number of non-terminal stellar eruptions. Though the progenitor class of these eruptions includes the most luminous stars, little is known of the pre-supernova mechanics of massive stars in their most evolved state, thus motivating a census of possible progenitors. From surveys of evolved and unstable luminous star populations in nearby galaxies, we select a sample of yellow and red supergiant candidates in M31 and M33 for review of their spectral characteristics and spectral energy distributions. Since the position of intermediate and late-type supergiants on the color-magnitude diagram can be heavily contaminated by foreground dwarfs, we employ spectral classification and multi-band photometry from optical and near-infrared surveys to confirm membership. Based on spectroscopic evidence for mass loss and the presence of circumstellar dust in their SEDs, we find that $30-40\%$ of the yellow supergiants are likely in a post-red supergiant state. Comparison with evolutionary tracks shows that these mass-losing, post-RSGs have initial masses between $20-40\,M_{\odot}$. More than half of the observed red supergiants in M31 and M33 are producing dusty circumstellar ejecta. We also identify two new warm hypergiants in M31, J004621.05+421308.06 and J004051.59+403303.00, both of which are likely in a post-RSG state.
  
\end{abstract} 

\keywords{galaxies:individual(M31, M33) -- stars:massive -- supergiants}

\section{Introduction}\label{sec:intro}

For many decades, the standard model of stellar evolution for massive stars ($\ge$~9 $M_{\odot}$) was characterized as the progression from main sequence OB star to red supergiant (RSG) to terminal supernova (SN) explosion. We now know that the evolutionary paths of massive stars, as well as their terminal state, depends strongly on mass loss and their mass-loss histories. It was recognized some time ago that stars above some initial mass ($\sim40-50\,M_{\odot}$) do not evolve to the RSG stage \citep{humphreys_1979}; however, due to mass loss---possibly eruptive \citep{humphreys_1994}---these massive stars then return to hotter temperatures, perhaps becoming Luminous Blue Variables (LBVs) or Wolf-Rayet (WR) stars prior to their terminal state. Stellar interior models also show that as stars shed their outer layers, the mass fraction of the He core increases, and  when it exceeds $60-70\%$ of the total, the star will evolve to warmer temperatures \citep{giannone_1967}. 

Lower mass supergiants that enter the RSG stage will either end their lives as Type II-P SNe or in some cases evolve back to warmer temperatures before the terminal explosion.  \citet{smartt_2009} recently identified what he called ``the red supergiant problem''---the lack of Type II-P and Type II-L SN progenitors with initial masses greater than 18 $M_{\odot}$. RSGs between 18 and 30 M$_{\odot}$ would presumably end their lives in some other manner. They might migrate on the HR diagram to warmer temperatures before their terminal explosions.

The RSG stage is a well-established high mass-losing phase. Mass-loss rates can be anywhere from $10^{-6}~M_{\odot}$~yr$^{-1}$ in RSGs \citep{mauron_2011} to as high as $10^{-4}~M_{\odot}$~yr$^{-1}$ in extreme stars like VY CMa and in the warm hypergiants \citep{humphreys_2013}.  What fraction of the RSGs return to warmer temperatures, the physical characteristics of candidate post-RSGs, and their locations on the HR Diagram are thus crucial to our understanding the final stages of the majority of massive stars.

Due in part to their position on the HR Diagram (HRD), few post-RSGs are known. As yellow supergiants (YSGs), with spectral types from late A to K, they  occupy a transient state between the blue and red supergiants and may either be evolving from the main sequence to cooler temperatures, or back to warmer temperatures from the RSG stage. Both populations represent a relatively short transition state. In the Galaxy, the warm hypergiants, close to the upper luminosity boundary in the HRD with high mass-loss rates, enhanced abundances, and dusty circumstellar (CS) environments, are excellent candidates for post-RSG evolution.  These stars contrast with the intermediate-type yellow supergiants which have normal spectra in their long-wavelength spectral energy distributions (SEDs)---that is, no evidence for circumstellar dust or mass loss in their spectra. \citet{dejager_1998AR} has suggested that all of the mass-losing, high-luminosity F and G-type supergiants are in a post-RSG state.  The Galactic hypergiant IRC~+10420 has long been acknowledged as a post-RSG \citep{jones_1993,oud_1996}. With its complex CS environment, large infrared excess, high mass-loss rate, and mass-loss history, it is in many ways the best example \citep{humphreys_1997,humphreys_2002,oud_1998,shenoy_2016}. Others include HR 5171A and HR 8752 \citep{nieu_2012}. In M33, the peculiar Variable A, a high luminosity F-type hypergiant  \citep{humphreys_2006}, with its apparent transit in the HR Diagram  to cooler temperatures  due to a high mass-loss episode is another candidate for post-RSG evolution. \citet{humphreys_2013} (hereafter, Paper~I) has identified several additional warm hypergiants in M31 and M33 with dusty ejecta, strong stellar winds, and high mass-loss rates similar to their Galactic counterparts that are likely post-RSGs.

As part of our larger program on the luminous and variable stars in M31 and M33, in this paper we present a more comprehensive survey of the yellow and red supergiants. We use the presence of circumstellar dust in their long-wavelength SEDs and spectroscopic indicators of mass loss and winds to identify candidates for post-RSG evolution. We likewise use the presence of a large infrared excess in the SEDs of the red or M-type supergiants to identify those with high mass loss. One of the greatest observational challenges is to separate the member supergiants from the significant foreground population of yellow and red dwarfs and halo giants in the Galaxy. In the next section, we describe our target selection, foreground contamination, and observations. In \S\ref{sec:ysgs} we discuss the yellow supergiant population and our selection of the post-RSG candidates. The SEDs and the role of circumstellar dust on the luminosities of the red supergiants are presented in \S\ref{sec:rsgs}, and in \S\ref{sec:discussion} we present estimates of the mass loss for the dusty RSGs. In the last section, we discuss the resulting HRDs for the yellow and red supergiants, and compare the candidate post-RSG population with evolutionary track models.

\section{Sample Selection and Observations}\label{sec:observations}
\subsection{Target List}

Our targets were primarily selected from the published surveys of M31 and M33 for yellow and red supergiants \citep{drout_2009,massey_2009,drout_2012}. Their red and yellow candidates were all chosen from the Local Group Galaxies Survey (\LGGS; \citealt{massey_2007b}). Although their adopted magnitude limit and color range for the YSGs  ($V < 18.5$ and $0.4\leq\bv\leq1.4$) corresponds to that of F- and G-supergiants, the same color range will include a large fraction of foreground contamination from Galactic dwarfs and halo giants \citep{massey_2006,drout_2012}.

To establish membership for the yellow candidates, the \citet{drout_2009} M31 survey relied on radial-velocity measurements. However, as their Figure 10 illustrates, even restricting the candidates to the velocity range expected for M31 includes substantial foreground contamination. \citet{drout_2009} therefore used a relative velocity---the measured velocity compared to the expected velocity of the star at its position in M31---to establish probable membership. They identified 54 rank-1 (highly probable) and 66 rank-2 (likely) yellow supergiants in M31 from a sample of 2901 targets. The 96\% foreground contamination clearly demonstrates the difficulties of color and magnitude criteria for determining membership.

For the candidate YSGs in M33 \citep{drout_2012}, the authors again relied on the relative radial velocities, but also added a measurement of the strong luminosity sensitive \ion{O}{1} $\lambda$7774 blend in A and F-type supergiants, which greatly increased the probability that the stars were supergiant members. With these criteria, they identified 121 rank-1 YSGs and 14 rank-2 in M33.

Fortunately for the RSG candidates, the two-color \bv~vs.~\vr~diagram has been demonstrated as an effective metric for distinguishing red dwarfs and supergiants \citep{massey_2009,drout_2012}, from which the authors identify 437 RSG candidates in M31 and 408 in M33. For the M31 candidates, 124 had additional radial-velocity information for membership determination, and 16 were spectroscopically confirmed as M-type supergiants.  For M33, the 408 candidate RSGs from \citet{drout_2012} were reduced to 204 (189 rank-1, 15 rank-2) likely RSGs using radial-velocity criteria.

 In addition to the 120 and 135 YSG candidates from the Drout/Massey catalogs of M31 and M33, respectively, we include 18 confirmed YSGs from \citet{humphreys_2014} (hereafter, Paper~II), and seven warm hypergiants from Paper~I, 39 H$\alpha$ emission stars with intermediate colors from the survey by \citet{valeev_2010}, and seven H$\alpha$ emission sources from an unpublished survey by K. Weis (see Paper II).  With these catalogs, we assembled a final target list of 124 and 165 candidate YSGs (after cross-identification among the listed works) for spectroscopy from M31 and M33. We did not obtain follow-up spectroscopy of the RSG candidates; our discussion of them instead relies on published photometry and analysis of their SEDs for circumstellar dust (\S\ref{sec:rsgs}).

\subsection{Observations}\label{sec:obs}

Our spectra of the YSG candidates were obtained with the Hectospec Multi-Object Spectrograph \citep{fabricant_1998,fabricant_2005} on the MMT at Mount Hopkins over several observing sessions in 2013, 2014, and 2015. The Hectospec has a 1\degr\ field of view with 300 fibers each subtending 1\farcs5 on the sky.  We used the 600 line mm$^{-1}$ grating with a tilt of 4800\AA\ yielding $\approx2500$\AA\ coverage with 0.54\AA\ pixel$^{-1}$ resolution and $R$ of $\sim2000$. The same grating with a tilt of 6800\AA\ was used for the red spectra with similar coverage and resolution and $R$ of $\sim3600$. A total integration time for each field was 90 minutes for the red and 120 minutes for the blue. The journal of observations is in Table~\ref{tab:one}.

Due to the large angular size of M31 on the sky, observations of this galaxy were split across two fields, labeled A and B in Table~\ref{tab:one}, centered at 00:43:36.5 +41:32:54.6 and 00:41:15.9 +40:40:31.2, respectively.  Weather conditions at Mount Hopkins during the 2015 season prevented observations on the last set of supergiant candidates in the red filter setting.

\begin{deluxetable}{llcc}
  \tablecaption{Journal of Observations\label{tab:one}}

  \tabletypesize{\footnotesize}
  \tablecolumns{4}
  \tablewidth{0pt}
  \tablenum{1}
  \tablehead{
    \colhead{Target}    & 
    \colhead{Date}      &
    \colhead{Exp. Time} &
    \colhead{Grating, Tilt} \\
    \colhead{}          &
    \colhead{(UT)}      &
    \colhead{(minutes)} &
    \colhead{}
  }
  \startdata
  M31A-Blue & 2013 Sep 25 & 120 & 600l, 4800\AA \\
  M31A-Red & 2013 Sep 26 & 90 & 600l, 6800\AA \\
  M31B-Blue & 2013 Oct 12 & 120 & 600l, 4800\AA \\
  M31B-Red & 2013 Oct 9 & 90 & 600l, 6800\AA \\
  M33-Blue & 2013 Oct 7 & 120 & 600l, 4800\AA \\
  M33-Red & 2013 Oct 7 & 90 & 600l, 6800\AA \\
  M33-Blue & 2014 Nov 29 & 120 & 600l, 4800\AA \\
  M33-Red & 2014 Nov 16 & 90 & 600l, 6800\AA \\
  M31A-Blue & 2015 Sep 20 & 120 & 600l, 4800\AA \\
  M31A-Red & 2015 Sep 20 & 90 & 600l, 6800\AA \\
  M31B-Blue & 2015 Sep 20 & 120 & 600l, 4800\AA \\
  \enddata
\end{deluxetable}

The spectra were reduced using an exportable version of the CfA/SAO SPECROAD package for Hectospec data.\footnote{\scriptsize{External SPECROAD was developed at UMN by Juan Cabanela for use on Linux or MacOS X systems outside of the CfA. It is available online at \url{http://astronomy.mnstate.edu/cabanela/research/ESPECROAD/}.}}  The spectra were bias subtracted, flat-fielded, and wavelength calibrated.  Due to crowding, sky subtraction was performed using pre-selected sky fibers off the field of each galaxy. These sky fiber positions were chosen from H$\alpha$ maps in regions where nebular contamination would be minimized. Flux calibration was done in IRAF using standard stars Feige 34 and 66 from Hectospec observations during the 2013-2015 seasons.

In M31, we obtained spectra for 113 of the 120 YSG candidates from \cite{drout_2009} plus follow-up spectra for the 10 previously confirmed warm supergiant and hypergiant stars from Papers I and II (6 of which were cross-listed in the Drout catalog) for a total of 117 spectra. In M33, 71 of the 135 YSG candidates from \cite{drout_2012} were observed, plus 14 confirmed supergiants from Papers I and II (4 cross-listed in the Drout catalog), and 37 H$\alpha$-emission sources from \citet{valeev_2010} (15 cross-listed in the Drout and Humphreys catalogs) for a total of 103 spectra.  For all of these sources, as well as the remaining YSG candidates from the Drout catalogs for which we did not observe spectra, we obtain photometry from published catalogs, discussed in \S\ref{sec:ysgphot}.

\section{The Intermediate-Type or Yellow Supergiants}\label{sec:ysgs}

\subsection{Spectral Characteristics}\label{sec:yspty}
Our primary goal for the spectroscopy of the YSG candidates is to search for evidence of mass loss and winds from emission lines and P Cygni profiles when present. Since foreground contamination is an obstacle for identifying yellow and red supergiant members in external galaxies, the same spectra can be used for spectral and luminosity classification.  We refine the classification of the YSG candidates with established luminosity and spectral type indicators in the blue and red spectra.

The blends of \ion{Ti}{2} and \ion{Fe}{2} at $\lambda\lambda4172$-8 and $\lambda\lambda4395$-4400 are strong luminosity criteria in the blue when compared against \ion{Fe}{1} lines that show little luminosity sensitivity such as $\lambda4046$ and $\lambda4271$. The \ion{O}{1}~$\lambda7774$ triplet in the red spectra is also a particularly strong luminosity indicator in A- to F-type supergiants. This feature is present in all of the candidate YSGs from M33, since \citet{drout_2012} identified all probable members based on this criterion.

The \ion{Sr}{2}~$\lambda\lambda4077, 4216$ lines are especially useful for temperature classifications of the yellow supergiants.  Comparing the relative strength of \ion{Sr}{2}~$\lambda4077$ to the nearby H$\delta$ feature, and the relative strengths of the \ion{Fe}{2}~$\lambda$4233 and \ion{Ca}{1}~$\lambda$4226 lines, for example, provide a quick diagnostic for all F-type supergiants.

Later type supergiants (G- type) are identified by the growth of the G-band,
a wide absorption feature around $\lambda4300$\,\AA\ due to CH. Luminosity criteria for G-type stars is similar to that in F supergiants. The \ion{Mg}{1} triplet $\lambda\lambda5167, 72, 83$ is strong in the later type dwarfs and allows for filtering foreground contaminants from our sample.

In M31, we identify 75 yellow supergiants, including the previously-confirmed stars from Papers I and II.  Seventy of the 113 observed stars from \citet{drout_2009} are confirmed YSGs, and 42 are foreground dwarfs or subgiants.  Therefore, the \citet{drout_2009} M31 catalog was $\sim$35\% contaminated by foreground stars. The remaining 8 rank-1/rank-2 candidates \cite{drout_2009} for which we did not obtain spectra are analyzed in \S\ref{sec:ysgsed} for evidence of mass loss in their SEDs along with the confirmed YSGs. We identify 86 yellow supergiants in M33, which also includes the warm supergiants and hypergiants discussed in Papers I and II. Sixty-two of the 71 observed candidates from \citet{drout_2012} are spectroscopically confirmed as yellow supergiants. The remaining 9 observed sources were identified as foreground dwarfs. Thus, the M33 catalog was only $\sim$7\% contaminated by foreground stars. Since their M33 survey used the luminosity-sensitive \ion{O}{1}~$\lambda$7774 line in addition to relative velocities, the cleaner sample is not surprising.

Twelve YSGs in M31---including the hypergiants M31-004322.50, M31-004444.52, M31-004522.58, and hypergiant candidate J004621.08+421308.2 (see \S\ref{sec:hyper})---and 18 in M33 (including hypergiants B324, Var A, N093351, and N125093) exhibit spectroscopic evidence for stellar winds. The notable spectral features include P Cygni profiles in the hydrogen emission lines, broad wings in H$\alpha$ or H$\beta$ emission indicative of Thomson scattering, and [\ion{Ca}{2}]/\ion{Ca}{2} triplet emission. Example spectra are shown in Figure~\ref{fig:one} highlighting two stars with H$\alpha$ emission indicative of stellar winds and circumstellar outflows.

We find that approximately 17\% of the observed YSGs in M31 and 21\% in M33 demonstrate evidence for mass loss in their spectra.  We discuss the evidence for circumstellar dust ejecta in their SEDs in \S\ref{sec:ysgsed}.  Representative A- and F-type supergiants from both galaxies are illustrated in Figure~\ref{fig:two}.

\figOne
\figTwo

Table~\ref{tab:two} is a list of the confirmed YSGs in both galaxies, 75 in M31 and 86 in M33, with their spectral types. Notes to the table include comments on the evidence for winds and mass loss in their spectra and references to cross-identified objects. The rank is included for stars from the Drout surveys. Table~\ref{tab:appendix} in the Appendix lists all of the foreground stars---confirmed dwarf and subgiant stars.\footnote{\scriptsize{At the time of this writing, \cite{massey_2016} have posted a spectroscopic survey of supergiants in M31 and M33 to the arXiv. Of the 75 YSGs in M31 identified here, 40 had consistent spectral types in \cite{massey_2016}. Similarly, of the 86 YSGs identified in M33, 22 were given spectral types in their work. Differences in spectral classification are typically within the same spectral type, e.g. F5 vs. F8.}}

Reduced spectra for the confirmed YSGs and foreground stars observed from 2013 to 2015 can be found at \url{http://etacar.umn.edu/LuminousStars/M31M33/}.

\begin{deluxetable}{lcclllc}
  \tablecaption{Spectroscopically-Confirmed YSGs\label{tab:two}}
  \tabletypesize{\scriptsize}
  \tablenum{2}
  \tablecolumns{6}
  \tablewidth{0pt}
  \setlength{\tabcolsep}{0.05in}
  \tablehead{\colhead{Star Name} & \colhead{RA} & \colhead{DEC} & \colhead{Sp Type} & \colhead{Notes} & \colhead{Alt Desig/Ref\tablenotemark{a}} & \colhead{Rank\tablenotemark{b}}}
  \startdata
  \bf{M31} & & & & & & \\
  M31-004247.30 & J004247.30 & +414451.0 & F5 &  & Paper II & 2 \\
  M31-004322.50 & J004322.50 & +413940.9 & A8-F0 & warm hypergiant & Paper I &  \\
  M31-004337.16 & J004337.16 & +412151.0 & F8 &  & Paper II & 2 \\
  M31-004350.50 & J004350.50 & +414611.4 & A5 & P Cyg H em & Paper II & 2 \\
  & J004410.62 & +411759.7 & F2 &  &  & 2 \\
  M31-004424.21 & J004424.21 & +412116.0 & F5 &  & Paper II & 2 \\
  & J004427.76 & +412209.8 & F5 & neb em &  & 2 \\
  & J004428.99 & +412010.7 & F0 &  &  & 2 \\
  \tableline
  \bf{M33} & & & & & & \\
  M33C-4640 & J013303.09 & +303101.8 & A0-2 & weak He I, Fe II em & Paper II &  \\
  & J013303.40 & +303051.2 & F5 & neb em & V-021266  & 1 \\
  & J013303.60 & +302903.4 & F8-G0 & G-band &  & 1 \\
  & J013311.70 & +302258.9 & F0-2 &  & V-028576 &  \\
  & J013410.61 & +302600.5 & F5-8 &  & V-119710 &  \\
  M33-013442.14 & J013442.14 & +303216.0 & F8 &  & Paper II  & 1 \\
  & J013446.93 & +305426.5 & A2 &  &  & 1 \\
  \enddata
  \tablenotetext{a}{V- or N prefix indicates the source identification is from \citet{valeev_2010}.}
  \tablenotetext{b}{Ranks from \citet{drout_2009,drout_2012} specify if the source was a (1) ``highly likely'' or (2) ``possible'' supergiant.}
  \tablenotetext{\,}{(This table is available in its entirety in a machine-readable form in the online journal. A portion is shown here for guidance regarding its form and content.)}
\end{deluxetable}

\subsection{Multi-Wavelength Photometry}\label{sec:ysgphot}
For each source in our target list, we cross-identify the visual photometry from  the \LGGS\ \citep{massey_2006} with the  near- and mid-infrared photomery from \tMASS\ \citep{skrutskie_2006} at $J$, $H$, and $K_s$,  the Spitzer/\IRAC\ surveys of M31 \citep{mould_2008} and M33 \citep{mcquinn_2007,thompson_2009} at 3.6, 4.5, 5.8 and 8~\micron, and \WISE\ \citep{wright_2010} at 3.4~(W1), 4.6~(W2), 12~(W3), and 22~(W4)~\micron. For cross-identification between the \tMASS\ and \IRAC\ coordinates, we use a search radius of 0\farcs5.

The \WISE\ satellite has angular resolutions of 6\farcs1, 6\farcs4, 6\farcs5, and 12\farcs0 in the four bands, which presents some issues for cross-identification in the crowded M31 and M33 fields.  We selected a 6\arcsec\ search radius for matching the \LGGS/\tMASS\ coordinates to \WISE, which is consistent with the FWHM of the \WISE\ PSF at 3.4~\micron\ \citep{wright_2010}. Since the longer-wavelength photometry has such a large beamsize, we recognize that some of our matched candidates may contain multiple sources or be contaminated by PAH emission. To mitigate this, the prime candidates for circumstellar dust (as characterized by infrared excess in the \WISE\ bands, \S\ref{sec:ysgsed}) were each checked visually in the \tMASS\ $K_s$-band images, and the photometry was rejected if the sources were likely composites. The resulting multi-wavelength photometry for the spectroscopically-confirmed yellow supergiants, as well as the YSG candidates, are summarized in Table~\ref{tab:three}.

\cite{mould_2008} and \cite{mcquinn_2007} additionally provide catalogs of infrared variable sources in M31 and M33, respectively.  We checked each source for variability against those catalogs, as well as the \textit{DIRECT} survey \citep{kaluzny_1998,macri_2001}, and find low-amplitude fluctuations ($< 0.1$ mag) in the \IRAC\ bands, most likely associated with Alpha Cygni variability. For M33 sources, we also check against the \cite{hartman_2006} optical survey and find similarly low-level flux variability in $g'$, $r'$, and $i'$ band observations. YSGs and YSG candidates that show variability in either the optical or infrared are indicated in Table~\ref{tab:three}.

\begin{deluxetable}{lccccccccccccccccccc}
  \tablecaption{Photometry of YSGs and YSG candidates\label{tab:three}}
  \tabletypesize{\scriptsize}
  \setlength{\tabcolsep}{0.05in}
  \tablecolumns{18}
  \tablewidth{0pt}
  \tablenum{3}

  \tablehead{\colhead{Star Name\tablenotemark{a}} & \colhead{U} & \colhead{B} & \colhead{V} & \colhead{R} & \colhead{I} & \colhead{J} & \colhead{H} & \colhead{K} & \colhead{3.6\micron\tablenotemark{b}} & \colhead{4.5\micron} & \colhead{5.8\micron} & \colhead{8.0\micron} & \colhead{3.4\micron\tablenotemark{c}} & \colhead{4.6\micron} & \colhead{12\micron} & \colhead{22\micron} &\colhead{Var\tablenotemark{d}}}
  \startdata
  \bf{M31} & & & & & & & & & & & & & & & & & \\
  D-003907.59 & 18.0 & 17.6 & 16.7 & 16.3 & 15.8 & 15.3 & 14.8 & 14.9 & ... & ... & ... & ... & 14.6 & 14.5 & 12.1 & 9.1 &  \\
  D-004009.13 & 19.1 & 18.5 & 17.6 & 17.2 & 16.8 & 16.1 & 15.8 & 15.4 & 14.9 & 14.3 & 14.3 & 13.7 & 14.8 & 14.3 & 12.3 & 8.8 &  \\
  M31-004247.30 & 17.1 & 16.9 & 16.4 & 16.0 & 15.6 & 15.3 & 15.0 & 15.0 & ... & ... & ... & ... & 14.6 & 14.6 & 11.6 & 8.3 &  \\
  D-004255.16 & 19.1 & 18.7 & 17.8 & 17.3 & 16.8 & 16.1 & 15.7 & 15.5 & ... & ... & ... & ... & ... & ... & ... & ... &  \\
  M31-004337.16 & 18.6 & 17.2 & 17.0 & 16.6 & 16.1 & 15.8 & 15.4 & 15.5 & 14.4 & 14.5 & 13.6 & 12.7 & 14.6 & 14.1 & 11.4 & 8.6 & V \\
  \tableline
  \bf{M33} & & & & & & & & & & & & & & & & & \\
  D-013231.94 & 18.5 & 18.1 & 17.4 & 17.0 & 16.6 & ... & ... & ... & 15.3 & 15.1 & ... & ... & 14.4 & 14.1 & 9.4 & 6.5 &  \\
  Var A & 20.1 & 19.8 & 18.8 & 18.2 & 17.7 & ... & ... & ... & 13.3 & 12.2 & 11.4 & 10.2 & 13.2 & 12.0 & 8.8 & 7.4 & V \\
  M33C-4640 & 16.4 & 17.1 & 17.0 & 16.9 & 16.7 & ... & ... & ... & 16.2 & 16.6 & ... & ... & 16.6 & 15.9 & 12.8 & 8.9 &  \\
  D-013439.98 & 17.6 & 17.3 & 16.8 & 16.5 & 16.1 & 15.7 & 15.4 & 15.2 & 15.0 & 15.0 & ... & ... & 15.2 & 15.0 & 12.6 & 8.8 &  \\
  M33C-013442.14 & 18.4 & 18.2 & 17.3 & 16.9 & 16.4 & 16.0 & 15.2 & 14.6 & 13.7 & 13.2 & ... & 11.7 & 13.8 & 13.0 & 10.4 & 8.1 &  \\
  \enddata
  \tablenotetext{a}{D- indicates the source was listed in \citet{drout_2009,drout_2012} with the name specifying the RA coordinate of its \LGGS\ ID. M31- or M33C- indicate a star name given in Paper I or II. The shorthand naming convention is for ease of matching to other tables in this paper. The complete RA and DEC designations are provided in Table~\ref{tab:two} and in the electronic version.}
  \tablenotetext{b}{3.6, 4.5, 5.8, and 8.0~\micron\ photometry from Spitzer/\textit{IRAC}.}
  \tablenotetext{c}{3.4, 4.6, 12, and 22~\micron\ photometry from \textit{WISE}.}
  \tablenotetext{d}{Indicates the source was identified as variable in the \textit{IRAC} bands in \cite{mould_2008} for M31 or \cite{mcquinn_2007} for M33, or variable in the optical from the \textit{DIRECT} survey \citep{kaluzny_1998,macri_2001} and from \cite{hartman_2006} for M33.}
  \tablenotetext{\,}{(This table is available in its entirety in a machine-readable form in the online journal. A portion is shown here for guidance regarding its form and content.)}
\end{deluxetable}

\subsection{Extinction Correction and the Spectral Energy Distributions}\label{sec:ysgsed}
To determine whether the yellow supergiants have excess free-free emission in the near-infrared ($1-2$~\micron) due to stellar winds and/or an excess at longer wavelengths due to circumstellar dust, we must first correct the SEDs for interstellar extinction. Many of these targets are likely embedded in their own circumstellar ejecta or warm circumstellar dust. Additionally, we have noticed in our previous work that the extinction can vary considerably across the face of these galaxies, especially in M31.  \citet{drout_2009} assumed a fixed $E(B-V) = 0.13$ reddening law for all YSGs in M31, and \citet{drout_2012} similarly adopted $E(B-V) = 0.12$ for sources in M33. We instead proceed more conservatively and calculate the extinction for each source individually.

For those stars with spectral types, we compare the observed \bv~color to the intrinsic colors of supergiants from \citet{flower_1977} and calculate $A_V$ from the standard extinction curves \citep{cardelli_1989} with R $=$ 3.2. This procedure is uncertain for stars with strong  emission lines in their spectra, so we also estimate the visual extinction using two other methods: the reddening-free $Q$-method \citep{hiltner_1956,johnson_1958} for nearby OB-type stars in the \LGGS\ within $2-3\arcsec$ of each target, assuming that their $UBV$ colors are normal, and the relation between the neutral hydrogen column density ($N_H$) and the color excess, $E_{B-V}$ \citep{savage_1972,knapp_1973}\footnote{\scriptsize{for R $=$ 3.2, $N_H/A_V = 1.56\times10^{21}\;\text{atoms}\;\text{cm}^{-2}\;\text{mag}^{-1}$\,\,\,[\citet{rachford_2009}].}}

We measure $N_{H}$ from the recent \ion{H}{1} surveys of M31 \citep{braun_2009} and M33 \citep{gratier_2010}.  Since we do not know the exact location of the stars along the line-of-sight with respect to the neutral hydrogen, we follow Paper~II and define the total $A_V$ as the foreground $A_V$ ($\approx$ 0.3 mag) plus half of the measured $N_{H}$. Since the \ion{H}{1} surveys have spatial resolutions of 30\arcsec\ and 17\arcsec\ for M31 and M33, respectively, we favor the extinction estimates from the two other methods when available. We use the $Q$-method and $N_H$ measurements for the supergiant candidates for which we did not obtain spectra. The results from these different methods, the adopted $A_V$, and the resulting extinction-corrected $M_{V}$ are summarized in Table~\ref{tab:four}.

For the spectroscopically-confirmed YSGs with known spectral types, we calculate the bolometric luminosities by applying bolometric corrections from \citet{flower_1996} to $M_{V}$. Bolometric corrections for stars in this temperature range are small, typically $\le |0.2|$ mag. For the candidate supergiants without spectra, we integrate the SED from the optical to the \tMASS\ $K_s$ band (2.2~\micron).  If the SEDs show an infrared excess, and thus evidence of circumstellar dust, we integrate the SED out to the \IRAC\ 8~\micron\ band and/or the 22~\micron\ \WISE\ band if available and if not obviously contaminated by nebulosity in the beam. Those sources are indicated with an asterisk in Table~\ref{tab:four}.

Figures~\ref{fig:three} and \ref{fig:four} are example SEDs from supergiants in M31 and M33.  The observed visual, \tMASS, and \IRAC\ magnitudes are shown as filled circles, and the \WISE\ data as open circles. The optical and \tMASS\ extinction-corrected photometry are shown as grey boxes.  We fit a blackbody to the extinction-corrected optical photometry to model the contribution from the central star. If the flux in the near-infrared \tMASS\ and \IRAC\ bands exceeds the expected Rayleigh-Jeans tail of the stellar component, we identify this as an infrared excess. Many of the supergiants in our sample, show the characteristic upturn redward of 8~\micron\ due to PAH emission \citep{draine_2007}.  Due to the large beamsize of \WISE, it is likely that some sources are contaminated by \ion{H}{2} region PAH emission in the mid-infrared.  However, if a source also has an apparent excess in the near-infrared \tMASS\ and \IRAC\ bands, the infrared photometry provides evidence of mass loss.  An infrared excess in the $1-2$~\micron\ \tMASS\ bands is characteristic of free-free emission in stellar winds, with the 3.6 to 8~\micron\ \IRAC\ data providing evidence for warm CS dust. Free-free emission is generally identified as constant F$_{\nu}$ in the near-infrared, often extending out to 5~\micron\ (see Figure~\ref{fig:six}). The \IRAC\ photometry can be used to estimate the mass of the dusty circumstellar material (see \S\ref{sec:masslossdust}). We note that the data provided in the broadband visual (\LGGS), near- (\tMASS) and mid-infrared (\IRAC\ and \WISE) photometry were not all observed simultaneously. The resulting SEDs, then, do not represent a single snapshot in time.

\begin{deluxetable}{llcccccl}
  \tablecaption{Extinction and Luminosities of YSGs and YSG Candidates\label{tab:four}}
  \tabletypesize{\scriptsize}
  \setlength{\tabcolsep}{0.05in}
  \tablecolumns{8}
  \tablewidth{0pt}
  \tablenum{4}
  
  \tablehead{\colhead{Star Name} & \colhead{Sp Type} & \colhead{$A_V$ (colors)} & \colhead{$A_V$ (stars)} & \colhead{$A_V$ ($N_H$)} & \colhead{Adopted $A_V$} & \colhead{$M_V$} & \colhead{$M_{Bol}$\tablenotemark{*}}}
  \startdata
  \bf{M31} & & & & & & & \\
  D-003926.72 & A2 & 0.7 & 0.7 & 1.4 & 0.7 & -7.1 & -7.3 \\
  D-003936.96 & ... & ... & 0.9 & 0.5 & 0.9 & -7.2 & -7.5* \\
  D-003948.85 & F2-5 & 1.2 & ... & 1.0 & 1.2 & -8.3 & -8.4 \\
  M31-004247.30 & F5 & 0.6 & ... & 0.9 & 0.6 & -8.7 & -8.9* \\
  M31-004522.58 & A2 & 0.4 & 1.4 & 1.1 & 0.4 & -6.4 & -7.2* \\
  \tableline
  \bf{M33} & & & & & & & \\
  M33C-4640 & A0-2 & 0.6 & 0.6 & 0.6 & 0.6 & -8.1 & -8.3 \\
  N045901 & F5 & 1.2 & 0.9 & 0.5 & 1.2 & -8.5 & -8.4* \\
  V071501 & A5 & 0.6 & ... & ... & 0.6 & -7.0 & -7.0 \\
  N125093 & F0-2 & ... & ... & 0.8 & 0.8 & -8.8 & -8.9* \\
  D-013439.73 & A5-8 & 0.6 & 0.9 & 0.6 & 0.6 & -8.0 & -8.0 \\
  \enddata
  \tablenotetext{*}{\,indicates the presence of an IR excess, and thus $M_{Bol}$ was calculated by integrating the SED out to the mid-infrared.}
  \tablenotetext{\,}{(This table is available in its entirety in a machine-readable form in the online journal. A portion is shown here for guidance regarding its form and content.)}

\end{deluxetable}

\figThree
\figFour

Twenty-six YSG candidates have indicators for free-free emission in the near-IR \tMASS\ photometry and/or CS dust emission in the mid-IR \IRAC\ or \WISE\ bands.  D-004009.13, as its SED in Figure~\ref{fig:three} shows, likely has both nebular contamination and dust. Combining both the spectroscopic and photometric data, we find a total of 32 sources in M31 with evidence for mass loss either from the stellar wind features in their spectra or free-free/CS dust emission in their SEDs. Six show evidence for both; the warm hypergiants: M31-004322.50, M31-004444.52, M31-004522.58, the new hypergiant J004621.08+421308.2 (see \S\ref{sec:hyper}), and two F-type supergiants: M31-004424.21, M31-004518.76.  We do not have spectra for two of the sources with evidence of free-free emission, D-003745.26\footnote{\scriptsize{This source is identified as an F5 supergiant by \cite{massey_2016}.  D-003936.96 and the six others for which we did not observe spectra remain unclassified.}}  and D-003936.96, so we cannot confirm membership in M31. Therefore, of the 75 confirmed YSGs in M31,  30 (or 40\%) are likely post-RSG candidates, plus two sources that require follow-up spectroscopy to confirm supergiant status.  Five sources (D-003711.98, D-003725.57, D-003907.59, D-004102.78, D-004118.69) are likely contaminated with nebular PAH emission from nearby \ion{H}{2} regions.

In M33, 22 stars show evidence for free-free emission and/or CS dust emission in their SEDs. Combining the spectroscopic and photometric data indicators, we find a total of 30 sources in M33 with evidence for mass loss. Eight have both the spectroscopic stellar wind and circumstellar dust features; the warm hypergiants: Var A, M33-013442.14, N093351, N125093, and the supergiants: V002627, D-013233.85, V021266, V130270, V104958. Three of the sources have not been spectroscopically confirmed as members of M33 (D-013345.50, D-013349.85, D-013358.05)\footnote{\scriptsize{These three sources are also in the \cite{massey_2016} catalog, but without spectral types.}}.  Therefore, of the 86 confirmed YSGs in M33, we identify 27 (or $\sim$31\%) as likely post-RSG candidates, plus three sources requiring follow-up spectroscopy. Thirty-three sources show evidence for nebular PAH contamination.

In Table~\ref{tab:five} we list all the stars, both confirmed and unconfirmed YSGs, that show at least one indicator of circumstellar ejecta: spectroscopic evidence of a stellar wind, free-free emission and/or thermal dust emission in the SED. If the stellar spectrum contains nebular emission markers such as [\ion{O}{3}], we indicate in the table that the source is likely contaminated with nebular emission. Estimates of total mass lost are discussed in \S\ref{sec:masslossdust}.

\begin{deluxetable}{llcclcc}
  \tablecaption{YSG and YSG Candidates with Evidence for Stellar Winds and CS Dust\label{tab:five}}
  \tabletypesize{\scriptsize}
  \setlength{\tabcolsep}{0.05in}
  \tablecolumns{6}
  \tablewidth{0pt}
  \tablenum{5}
  \tablehead{\colhead{Star Name} & \colhead{Sp Type} & \colhead{Wind} & \colhead{IR excess} & \colhead{Comments} & \colhead{Mass Lost ($M_{\odot}$)\tablenotemark{a}}}
  \startdata
  \bf{M31} & & & & & \\
  D-004009.13 & F2-5 & ... & yes & CS dust & $0.07^{\pm0.01}\times 10^{-2}$ \\
  M31-004322.50 & A8-F0 & yes & yes & CS dust & $0.08^{\pm0.01}\times 10^{-2}$ \\
  M31-004337.16 & F8 & ... & yes & CS dust & $0.17^{\pm0.02}\times 10^{-2}$ \\
  M31-004424.21 & F5 & yes & yes & CS dust & $1.42^{\pm0.18}\times 10^{-2}$ \\
  M31-004444.52 & F0 & yes & yes & CS dust & $1.74^{\pm0.22}\times 10^{-2}$ \\
  \tableline
  \bf{M33} & & & & & \\
  D-013231.94 & F2 & yes & yes & CS dust & $0.76^{\pm0.09}\times 10^{-2}$ \\
  Var A & F8 & yes & yes & CS dust & $2.36^{\pm0.28}\times 10^{-2}$ \\
  D-013349.86 & F8 & ... & yes & CS dust & $0.87^{\pm0.10}\times 10^{-2}$ \\
  M33-013357.73 & A0 & yes & yes & CS nebula, H II PAH & ... \\
  N093351 & F0 & yes & ... &  & $2.06^{\pm0.24}\times 10^{-2}$ \\
  \enddata
  \tablenotetext{a}{Total mass lost through circumstellar ejecta estimated from \IRAC/\WISE\ photometry. See \S\ref{sec:masslossdust}.}
  \tablenotetext{\,}{(This table is available in its entirety in a machine-readable form in the online journal. A portion is shown here for guidance regarding its form and content.)}
\end{deluxetable}

\subsection{Two New Hypergiants in M31}\label{sec:hyper}
J004621.05+421308.06 was considered a candidate LBV by \citet{massey_2007c} and \citet{king_1998}. Due to its position in the northern arm of M31, it was outside both of our two Hectospec fields for M31. Consequently, we obtained a long-slit spectrum of this target on 22 November 2014 with the MODS1 spectrograph on the Large Binocular Telescope. The instrument setup and observing procedure are described in Paper~I.  Its blue and red spectra are shown in Figure~\ref{fig:five}.  J004621.05+421308.06 has the absorption-line spectrum of a late A-type supergiant with strong Balmer emission lines with deep P Cygni profiles and the broad wings characteristic of Thomson scattering. The outflow velocity is -223~km~s$^{-1}$, measured from three P Cygni absorption minima. Numerous \ion{Fe}{2} and [\ion{Fe}{2}] emission lines are present. Like the warm hypergiants discussed in Paper~I, its red spectrum shows the \ion{Ca}{2} and [\ion{Ca}{2}] line emission indicative of a low-density circumstellar nebula. The \ion{O}{1}~$\lambda$8846 line is also in emission. The SED shown in Figure~\ref{fig:six} reveals a prominent circumstellar dust envelope in the near- and mid-infrared not observed in the LBVs (Paper~II). These properties are shared with the warm hypergiants and probable post-red supergiants experiencing high mass loss.  Their photospheres are not due to the cool dense winds formed by an LBV in eruption, but represent the stellar surface. We therefore suggest that J004621.05+421308.06 belongs with the class of warm hypergiants.

J004051.59+403303.00 has been described by \citet{massey_2007c} and by \citet{sholukhova_2015} as a candidate LBV. Our blue and red spectra from 2013 shown in Figure~\ref{fig:five} are similar to the blue spectrum published by \citet{massey_2006} and the blue and red spectra in \citet{sholukhova_2015} suggesting little spectroscopic change in the last 10 years.  The spectra show prominent P Cygni profiles in the Balmer lines with broad wings and in the \ion{Fe}{2} multiplet 42 lines. The mean outflow velocity measured from the absorption minimum in six P Cygni profiles is -152~km~s$^{-1}$ similar to the LBVs and hypergiants (Papers~I and II).  There are no other \ion{Fe}{2} or [\ion{Fe}{2}] emission lines. The relative strengths of the \ion{Mg}{2} $\lambda$4481 and \ion{He}{1} $\lambda$4471 lines suggest an early A-type supergiant. Its spectrum is similar to the warm hypergiant M31-004444.52 in Paper~I, but it does not have the [\ion{Ca}{2}] emission lines in the red. It also resembles J004526.62+415006.3 in 2010 which was later shown to be an LBV entering its maximum light or dense-wind stage \citep{sholukhova_2015,humphreys_2015}. Therefore, based on this spectrum, its nature is somewhat ambiguous. Its SED in Figure~\ref{fig:six} shows an excess in the near-infrared due to free-free emission,  as evidenced by constant flux (F$_{\nu}$) out to 5~\micron. The \WISE\ photometry at 12 and 22~\micron\ may be due to circumstellar dust from silicate emission, but is more likely contaminated by PAH emission from a nearby \ion{H}{2} region and nebulosity.

Thus, J004051.59+403303.00 may be a mass-losing post-red supergiant like several of the stars discussed in this paper or a candidate LBV. Future spectroscopic and photometric variability will be necessary to confirm that it is an LBV, but even so, given its luminosity, it is very likely in a post-RSG state similar to the less-luminous LBVs.

\figFive
\figSix 

\section{Red Supergiants}\label{sec:rsgs}
With significant mass loss, red supergiants can evolve back to warmer temperatures, we examine the SEDs of the RSGs to identify what fraction of these cool supergiants are in a mass-losing state and determine their positions on the HRD.  Additionally, we can roughly estimate the total mass lost through circumstellar ejecta from the stars' infrared photometry.  The RSGs currently experiencing episodes of high mass loss may eventually evolve to become post-RSG warm supergiants, LBVs, or WR stars.

We cross-identify the visual photometry for the RSGs from the \LGGS\  with \tMASS, \IRAC, and \WISE. The multi-wavelength photometry is summarized in Table~\ref{tab:six} for the 437 RSG candidates in M31 \citep{massey_2009} and the 204 (189 rank-1 plus 15 rank-2) in M33 \citep{drout_2012}.

\begin{deluxetable}{lcccccccccccccccccc}
  \tablecaption{Photometry of Candidate RSGs\label{tab:six}}
  \tabletypesize{\scriptsize}
  \setlength{\tabcolsep}{0.05in}
  \tablecolumns{17}
  \tablewidth{0pt}
  \tablenum{6}

  \tablehead{\colhead{Star Name\tablenotemark{a}} & \colhead{U} & \colhead{B} & \colhead{V} & \colhead{R} & \colhead{I} & \colhead{J} & \colhead{H} & \colhead{K} & \colhead{3.6\micron\tablenotemark{b}} & \colhead{4.5\micron} & \colhead{5.8\micron} & \colhead{8.0\micron} & \colhead{3.4\micron\tablenotemark{c}} & \colhead{4.6\micron} & \colhead{12\micron} & \colhead{22\micron}}
  \startdata
  \bf{M31} & & & & & & & & & & & & & & & & & \\
  M-003703.64 & ... & 21.6 & 19.9 & 19.0 & 17.7 & 16.6 & 15.7 & 15.7 & ... & ... & ... & ... & 15.5 & 15.7 & 12.9 & 9.4 \\
  M-003723.56 & 22.1 & 20.8 & 18.9 & 17.9 & 16.9 & 15.7 & 14.9 & 14.8 & ... & ... & ... & ... & 14.5 & 14.7 & 12.4 & 9.3 \\
  M-003724.48 & ... & 21.7 & 19.6 & 18.5 & 17.5 & 16.2 & 15.0 & 15.1 & 16.3 & 16.2 & 13.8 & 11.8 & 14.8 & 15.1 & 12.7 & 9.4 \\
  M-004120.25 & 23.1 & 20.9 & 18.9 & 17.6 & 16.3 & 14.9 & 14.0 & 13.7 & 13.0 & 13.2 & 12.9 & 12.4 & 13.4 & 13.6 & 12.7 & 9.2 \\
  M-004444.66 & 22.5 & 20.9 & 19.0 & 18.0 & 16.9 & 15.8 & 14.8 & 14.8 & 14.2 & 14.3 & 14.1 & 14.1 & 14.5 & 14.6 & 12.3 & 9.1 \\
  \tableline
  \bf{M33} & & & & & & & & & & & & & & & & & \\
  D-013217.79 & 20.5 & 19.6 & 18.4 & 17.8 & 17.2 & 16.6 & 15.7 & 15.9 & 15.8 & 15.8 & ... & ... & 15.8 & 16.0 & 12.9 & 8.8 \\
  D-013224.33 & 20.9 & 21.4 & 19.6 & 18.7 & 17.8 & 16.7 & 16.1 & 15.6 & 15.7 & 15.9 & ... & ... & 15.7 & 15.9 & 12.4 & 9.5 \\
  D-013312.26 & 19.1 & 17.6 & 16.0 & 15.1 & ... & 13.3 & 12.6 & 12.4 & 12.4 & 12.4 & ... & 12.1 & 12.3 & 12.3 & 12.0 & 8.7 \\
  D-013421.55 & ... & 21.2 & 19.3 & 18.1 & 16.8 & 15.4 & 14.6 & 14.4 & 14.2 & 14.3 & ... & 13.5 & 14.0 & 14.1 & 12.3 & 8.7 \\
  D-013502.06 & 21.0 & 19.9 & 18.5 & 17.9 & 17.3 & 16.5 & 16.0 & 16.6 & 15.8 & 15.8 & ... & ... & 15.9 & 16.0 & 12.8 & 9.0 \\  
  \enddata
  \tablenotetext{a}{D- indicates the source was listed in \citet{drout_2012} with the name specifying the RA coordinate of its \LGGS\ ID. M- indicates the source was listed in \citet{massey_2009}. The shorthand naming convention is for ease of matching to other tables in this paper. The complete RA and DEC designations are provided in the electronic version.}
  \tablenotetext{b}{3.6, 4.5, 5.8, and 8.0~\micron\ photometry from Spitzer/\textit{IRAC}.}
  \tablenotetext{c}{3.4, 4.6, 12, and 22~\micron\ photometry from \textit{WISE}.}
\end{deluxetable}

Since we lack spectral type information, we cannot correct for extinction using intrinsic colors. \citet{massey_2009} applied a constant $A_V = 1$ to the entire sample of M31 RSGs, and \citet{drout_2012} adopted a fixed reddening law of $E(B - V) = 0.12$ for all RSGs in M33. Here, we follow the methodology for the YSGs and estimate $A_V$ from the $Q$-method for nearby O and B stars and from the neutral hydrogen column density along the line-of-sight to each RSG as described in \S\ref{sec:ysgsed}.  Unfortunately, roughly 60\% of the RSG candidates lacked nearby O and B stars, so we are forced to adopt the less accurate extinction from the neutral hydrogen. The results from the two methods are summarized in Table~\ref{tab:seven}. Bolometric luminosities are calculated by integrating the optical through \tMASS\ $K_s$. Similar to the YSG sources, if the SEDs show an infrared excess in the \tMASS\ or \IRAC\ photometry of the RSG candidates, we integrate the SED out to the \IRAC\ 8~\micron\ band and/or the 22~\micron\ \WISE\ band if uncontaminated.  Those sources are indicated with an asterisk in Table~\ref{tab:seven}.  Several sources were found to have anomalous photometry in the optical or infrared, possibly due to crowding in the field or source mismatch from the \LGGS.  Some of these objects with unusually high bolometric luminosities may actually be foreground stars, but without spectra, we cannot confirm membership.  These stars are omitted from the HR diagrams in \S\ref{sec:HRdiag}. They are included in the catalogs for completeness and are indicated with a dagger in Table~\ref{tab:seven}.

\begin{deluxetable}{lccccl}
  \tablecaption{Extinction and Luminosities of Candidate RSGs\label{tab:seven}}
  \tabletypesize{\scriptsize}
  \setlength{\tabcolsep}{0.05in}
  \tablecolumns{6}
  \tablewidth{0pt}
  \tablenum{7}
  \tablehead{\colhead{Star Name} & \colhead{$A_V$ (stars)} & \colhead{$A_V$ ($N_H$)} & \colhead{Adopted $A_V$} & \colhead{$M_V$} & \colhead{$M_{Bol}$\tablenotemark{*}}}
  \startdata
  \bf{M31} & & & & & \\
  M-003739.41 & ... & 0.8 & 0.8 & -5.9 & -6.7 \\
  M-003739.88 & 0.9 & 0.7 & 0.9 & -6.7 & -8.2* \\
  M-003907.69 & 1.8 & 1.4 & 1.8 & -7.5 & -8.3 \\
  M-003907.98 & 1.6 & 1.4 & 1.6 & -7.4 & -8.0* \\
  M-004638.17 & 1.2 & 1.1 & 1.2 & -6.1 & -8.3* \\
  \tableline
  \bf{M33} & & & & & \\
  D-013339.28\tablenotemark{\dagger} & 0.7 & ... & 0.7 & -8.4 & -9.8* \\
  D-013340.80 & 0.8 & 0.6 & 0.8 & -6.1 & -7.0 \\
  D-013349.09 & 0.9 & 0.6 & 0.9 & -6.3 & -8.0* \\
  D-013349.99 & 1.9 & 1.0 & 1.9 & -7.1 & -8.2* \\
  D-013438.95 & 0.4 & 0.5 & 0.4 & -5.5 & -7.5* \\
  \enddata
  \tablenotetext{*}{\,indicates the presence of an IR excess, and thus $M_{Bol}$ was calculated by integrating the SED out to the mid-infrared.}
  \tablenotetext{\dagger}{Photometry of sources is anomalous. For some stars in crowded fields, there may be either a source mismatch between the optical and infrared or the photometry may be contaminated by multiple sources in the aperture.  M-004539.99, D-013312.26, and D-013401.88 (included in full table online) are likely foreground stars. Marked sources are omitted from the HR diagrams for above reasons.}
  \tablenotetext{\,}{(This table is available in its entirety in a machine-readable form in the online journal. A portion is shown here for guidance regarding its form and content.)}
\end{deluxetable}

Figures~\ref{fig:seven} and \ref{fig:eight} are example SEDs for RSGs in both galaxies.  The symbols follow the same pattern as Figure~\ref{fig:three}.  Since these stars are cooler, the peak of the optical thermal component from the star shifts redward in the SED.  This makes any infrared excess in the \tMASS\ bands less discernable than in the YSGs; however, the CS dust component at wavelengths longer than 3.6~\micron\ can still be readily distinguished in most of the RSG candidate sources. For this reason, we divide our SEDs into rankings.  Rank-1 SEDs have an infrared excess in the \IRAC\ and/or \WISE\ bands most probably due to CS dust emission. Rank-2 SEDs either have missing \IRAC\ photometry but show an IR excess in \WISE, or have an IR excess in the \IRAC\ bands but which is somewhat uncertain due to the characteristic PAH upturn in the \WISE\ bands and are thus possibly contaminated by nebulosity. Figure~\ref{fig:seven} demonstrates two rank-1 RSGs, with SEDs showing excess emission above the color-temperature fits to the optical data. The bottom panel of Figure~\ref{fig:eight} illustrates one of the more ambiguous sources in M33. D-013353.91 has a clear infrared excess at 8~\micron, while the infrared photometry D-013506.97 can be easily confused with nebulosity. We consider D-0133506.97 as a rank-2 mass-losing RSG candidate.

\figSeven
\figEight

Of the 437 RSG candidates in M31 from \cite{massey_2009}, 231 (129 rank-1 and 102 rank-2) show evidence for circumstellar dust emission in the mid-IR \IRAC\ or \WISE\ bands. Thus, $\sim$53\% of the candidate M31 RSGs have CS dust.\footnote{\scriptsize{Of the 231 candidate RSGs with evidence for mass loss, 152 have spectral classifications from \cite{massey_2016}.}} An additional 110 candidate RSGs are likely contaminated with nebular emission.

\begin{deluxetable}{lllccc}
  \tablecaption{RSG Candidates with Evidence for CS Dust\label{tab:eight}}
  \tablenum{8}
  \tabletypesize{\scriptsize}
  \setlength{\tabcolsep}{0.05in}
  \tablecolumns{5}
  \tablewidth{0pt}
  \tablehead{\colhead{Star Name} & \colhead{LGGS} & \colhead{Rank\tablenotemark{a}} & \colhead{Mass Lost ($M_{\odot}$)\tablenotemark{b}}}
  \startdata
  \bf{M31} & & & \\
  M-003930.30 & J003930.30+404353.4 & 1 & $0.13^{\pm0.02}\times 10^{-2}$ \\
  M-004024.52 & J004024.52+404444.8 & 1 & $0.05^{\pm0.01}\times 10^{-2}$ \\
  M-004036.08 & J004036.08+403823.1 & 1 & $1.07^{\pm0.13}\times 10^{-2}$ \\
  M-004031.00 & J004031.00+404311.1 & 1 & $0.39^{\pm0.05}\times 10^{-2}$ \\
  M-004304.62 & J004304.62+410348.4 & 1 & $0.51^{\pm0.06}\times 10^{-2}$ \\
  \tableline
  \bf{M33} & & & \\
  D-013354.32 & J013354.32+301724.6 & 1 & $0.16^{\pm0.02}\times 10^{-2}$ \\
  D-013401.88 & J013401.88+303858.3 & 1 & $1.48^{\pm0.18}\times 10^{-2}$ \\
  D-013416.75 & J013416.75+304518.5 & 2 & $0.10^{\pm0.12}\times 10^{-2}$ \\
  D-013454.31 & J013454.31+304109.8 & 1 & $0.37^{\pm0.04}\times 10^{-2}$ \\
  D-013459.81 & J013459.81+304156.9 & 2 & $0.14^{\pm0.02}\times 10^{-2}$ \\
  \enddata
  \tablenotetext{a}{Rank 1 indicates that an infrared excess in the SED is most probably due to CS dust emission. Rank 2 indicates that features in the SED are likely caused by thermal dust emission but may be due to PAH contamination.}
  \tablenotetext{b}{Total mass lost through circumstellar ejecta estimated from \IRAC/\WISE\ photometry. See \S\ref{sec:masslossdust}.}
  \tablenotetext{\,}{(This table is available in its entirety in a machine-readable form in the online journal. A portion is shown here for guidance regarding its form and content.)}
\end{deluxetable}

In M33, 126 of the 204 candidate RSGs from \cite{drout_2012}, have indicators for CS dust emission in the infrared. Again dividing the 126 sources with infrared excess into ranks from their SEDs, we find 53 rank-1 (highly probable) RSGs and 73 rank-2 (likely) candidates.\footnote{\scriptsize{Unlike the RSG candidates in M31, these sources in M33 did not have spectral classifications in \cite{massey_2016}.}}  Thus, $\sim$60\% of our RSG candidates in M33 have evidence for dusty ejecta. Forty-three sources show the PAH upturn in their SEDs and are likely contaminated with nebulosity.

Table~\ref{tab:eight} summarizes the results from both galaxies. We find that more than half of the RSG candidates in M31 and M33 exhibit evidence for mass loss. This high fraction is not surprising since M supergiants have been known for decades to have mass loss and dusty circumstellar ejecta \citep{woolf_1969}.  Our results are consistent with the $\sim$45\% found by \cite{mauron_2011} in the Milky Way and LMC using the \textit{IRAS} 60~\micron\ band. Since we have included our rank-2 SEDs in this census, the fractions reported here may represent an overestimate.

\section{Discussion}\label{sec:discussion}
\subsection{Circumstellar Dust and Mass Loss}\label{sec:masslossdust}
Thermal emission from dust appears in the mid-infrared \IRAC\ data from 3.6 to 8~\micron\ and is also present in the \WISE\ photometry at longer wavelengths. With some assumptions about the dust grain parameters, we can estimate the mass of the circumstellar material from the mid-infrared flux (see Paper~I):
\begin{equation*}
  M_{dust} = \frac{4D^2\rho\lambda F_\lambda}{3\left(\lambda Q_{\lambda}/a\right)B
    _{\lambda}\left(T\right)}
\end{equation*}
where $D$ is the distance to the source (here, the average distance to M31/M33), $F_\lambda$ is the mid-infrared flux, $a$ is the grain radius, $\rho$ is the grain density, $Q_\lambda$ is the absorption efficiency factor for silicate dust grains, and $B_\lambda\left(T\right)$ is the blackbody emission at temperature $T$.

For many of the YSG and RSG candidates with an infrared excess, the flux is fairly constant across the mid-infrared, which implies that the dust is emitting over a range of temperatures and distances around the central star. Using the \citet{suh_1999} prescription for silicate dust grains and mass loss around AGB stars, we assume dust grain size of $a = 0.1\,\micron$ at a density $\rho = 3$\;g\,cm$^{-3}$ and an average temperature of 350 K. The absorption (and emission) efficiency factor $Q_\lambda$ is maximized at the 9.8~\micron\ Si--O vibrational mode \citep{woolf_1969} and the 18~\micron\ O--Si--O bending mode \citep{treffers_1974}, so the flux at these wavelengths would be the ideal tracers of thermal dust emission.  Since we lack photometry precisely centered on the silicate features, we calculate the dust mass using the flux at 8~\micron, or at 12~\micron\ (W3) if no \IRAC\ data exists for the sources. We assume a nominal gas-to-dust ratio of 100, which allows for the calculation of the total mass lost in each source.

The results are summarized in Table~\ref{tab:five} for the YSGs and Table~\ref{tab:eight} for the RSGs, where the error is calculated as the standard error propagation on the average distance to M31/M33 ($<5\%$) and the photometric errors for measured flux by \IRAC/\WISE\ ($<3\%\;/<9\%$~\citealt{hora_2004,wright_2010}).  For both YSGs and RSGs in M31 and M33, we find a range of at least a factor of 10 for the mass of the circumstellar material. Most of the supergiants have shed $\sim10^{-3}-10^{-2}\,M_{\odot}$, which is consistent with Paper~I.  \cite{mauron_2011} apply the \cite{dejager_1988} mass-loss prescription to Galactic RSGs to calculate an average mass-loss rate of $\sim10^{-6}\,M_{\odot}\;yr^{-1}$ from IRAS 60~\micron\ flux. For our dusty RSGs, we can approximate a timescale probed by the \IRAC\ photometry and thus compare our total integrated mass loss to typical RSG mass-loss rates. If we assume an average dust condensation distance of $\sim$250 AU and an outflow velocity of 20 km/s, we estimate $\sim$100 years for the dust condensation time---a rough timescale for the dust we observe at 8~\micron. Considering that the circumstellar ejecta most likely contains dust over a range of temperatures ($\sim150-400$~K), as well as the possibility of episodic mass loss in the more massive RSGs, an average mass-loss rate of $10^{-5}-10^{-4}\,M_{\odot}\;yr^{-1}$ is consistent with the total mass-lost estimates of $10^{-3}-10^{-2}\,M_{\odot}$ over the dust condensation timescale.

For the RSG populations in both galaxies, we plot bolometric luminosity vs. total mass lost in Figure~\ref{fig:nine}.  The \cite{dejager_1988} formulation predicts an increasing mass-loss rate with luminosity, and we find a similar trend with total mass lost as traced by dust.  We separate the rank-1 RSGs, those candidates with clear indication of mass loss in their SEDs, from the rank-2 RSG candidates. The rank-2 sources yield ejecta masses on the lower end of the RSG sample.  These RSG candidates likely have circumstellar dust, but the infrared excess was not as obvious as in the rank-1 SEDs, thus the lower derived mass-loss estimate. Since the highest luminosity RSGs also have the highest mass loss, these dusty RSGs may evolve back to warmer temperatures to become the intermediate-type post-RSGs discussed in this paper.

\figNine

\subsection{HR Diagrams}\label{sec:HRdiag}
The HR Diagrams for the yellow and red supergiants populations are shown in Figures~\ref{fig:ten} and~\ref{fig:eleven}. The temperatures for the YSGs are derived from the $(B-V)_0$ colors using the transformations in \cite{flower_1996} for intermediate-type supergiants.  As described in \S\ref{sec:ysgsed}, their luminosities are calculated based on the bolometric corrections given in \cite{flower_1996} or by integrating the SED for those stars with emission line spectra or with circumstellar dust.

There are several temperature scales in the literature for red supergiants. Since we do not have spectral types for the RSGs, we adopt the temperatures from \citet{massey_2009} for the M31 RSGs and from \citet{drout_2012} for M33 simply for the purpose of placing them on the HR diagram to compare with the YSG population.  Their M31 temperature scale is based on a color--temperature relationship from $(V-K)_0$ from MARCS atmosphere models \citep{gustafsson_2008}, while their temperatures for the M33 stars are based on $(V-R)_0$ color transformations from \citealt{levesque_2006} in the LMC, which has a metallicity similar to M33. The bolometric luminosities for the RSGs are determined from integrating their SEDs (\S\ref{sec:rsgs}).

Stellar evolution tracks from non-rotating models from \citet{ekstrom_2012} are shown on the HR diagrams for ZAMS masses of 15, 25, and 40 $M_{\odot}$.  In both galaxies, the post-RSG candidates are preferentially more abundant at higher luminosities. Comparison with the evolutionary tracks suggests that most of the progenitor main-sequence stars have masses $\gtrsim$ 20 $M_{\odot}$. Likewise, the dusty RSGs dominate the higher luminosities. This is most obvious for the M33 population with a smaller sample. This is not surprising, as we know from Figure~\ref{fig:nine} that the mass lost in the RSGs correlates with luminosity.

We note the presence of several ``warm'' RSGs in both galaxies. These red supergiant candidates, with temperatures upwards of 4000K, fall in the temperature range of the yellow supergiants. The temperature scales are somewhat uncertain, and without spectra of these objects, we cannot confirm that some of them may actually be yellow supergiants.

Labeled sources in Figures~\ref{fig:ten} and~\ref{fig:eleven} are the warm hypergiants from Paper~I, as well as the two new hypergiant candidates, J004051.59+403303.00 and J004621.05+421308.06, discussed in \S\ref{sec:hyper}.

\figTen
\figEleven

\section{Conclusion}\label{sec:conclusion}

We identify 75 spectroscopically-confirmed yellow supergiants in M31, including the three warm hypergiants from Paper~I and 86 in M33 including the 14 previously-known YSGs from Papers~I and II. The majority have normal absorption-line spectra, but a significant fraction, 30 in M31 and 27 in M33, show evidence for mass loss via stellar winds and/or circumstellar dust in their SEDs. Since the RSG stage is a well-established high mass-losing state, we consider these stars to be excellent candidates for post-red supergiant evolution. Thus, about $30-40\%$ of the observed YSGs are likely in a post-RSG state. The post-RSG candidates are more common at luminosities above $\sim10^{5}\,L_{\odot}$. Most appear to have initial masses of $20-40\,M_{\odot}$, and may be the evolutionary descendants of the more massive red supergiants that do not explode as supernovae \citep{smartt_2009}. The eventual fate of these stars may be either as ``less-luminous'' LBVs or WR stars before their terminal explosion; however, in his most recent review, \cite{smartt_2015} argues for an upper limit of $\approx$ 18 $M_{\odot}$ of supernova progenitors, and that more massive stars collapse directly to black holes.

The less-luminous LBVs (M$_{Bol}$ $\approx$ 8 to -9.5 mag) have high L/M values of $\sim 0.5$, compared to the B- and A-type supergiants in the same part of the HR diagram. The most likely explanation is that the LBVs have shed a significant fraction of their mass in a previous state and are now close to their Eddington limit. Consequently, they have also been considered as evidence for post-RSG evolution \citep{humphreys_1994,vink_2012}, and would have passed through the YSG region of the HR diagram in their evolution to warmer temperatures. Thus, the mass-losing yellow supergiants may be thought of as the progenitor class of the less-luminous LBVs.

We identify two new warm hypergiant candidates in M31, J004621.05+421308.06 and J004051.59+403303.00. The spectra of both stars show strong P Cygni absorption profiles in the Balmer emission lines with broad Thomson scattering wings. J004621.05+421308.06 also has strong \ion{Ca}{2} and [\ion{Ca}{2}] emission indicative of a circumstellar nebula plus dusty circumstellar ejecta.  Both stars are very likely in a post-RSG state. J004051.59+403303.00 is also considered a candidate LBV. If so, it would be one of the less-luminous LBVs, but future spectroscopy and photometry is necessary for confirmation.

The red supergiant sample yielded 231 stars in M31 (53\%) and 126 in M33 (60\%) with observable dusty emission. Therefore, a large fraction of RSGs are in a mass-losing state.  Consistent with \cite{mauron_2011} and the de Jager prescription, we find that mass loss correlates with luminosity along the RSG branch.  The \IRAC\ 8~\micron\ band provides a reasonable estimate of the total dust mass lost over a timescale of about a century, and we estimate that the RSGs in both galaxies tend to have dusty ejecta on the order of $10^{-3}-10^{-2}\,M_{\odot}$ assuming a warm dust component of 350 K.  If more than 50\% of RSGs are indeed experiencing sufficient mass loss to produce CS dusty ejecta, a large fraction of stars along the red supergiant branch may evolve back towards the blue to become the warm post-RSG stars before their terminal state as supernovae or black holes.

We note that our target selection was derived from optical surveys.  Therefore, our survey of the most luminous stars in M31 and M33 does not include supergiant stars that may be obscured. Since the most luminous warm and cool supergiant populations are more likely to have the highest mass-loss rates, it is probable that those sources will be highly obscured in the optical by their own circumstellar ejecta. To complete the upper portion of the HR diagram requires a further search through the \IRAC\ data to find the brightest infrared sources.

\vspace{5mm}
Research by M. Gordon and R. Humphreys on massive stars is supported by the National Science Foundation AST-1109394.  We thank Perry Berlind and Michael Calkins at the MMT for their excellent support and operation of the Hectospec. This paper uses data from the MODS1 spectrograph built with funding from NSF grant AST-9987045 and the NSF Telescope System Instrumentation Program (TSIP), with additional funds from the Ohio Board of Regents and the Ohio State University Office of Research. This publication also makes use of data products from the Two Micron All Sky Survey, which is a joint project of the University of Massachusetts and the Infrared Processing and Analysis Center/California Institute of Technology, funded by the National Aeronautics and Space Administration and the National Science Foundation, and from the Wide-field Infrared Survey Explorer, which is a joint project of the University of California, Los Angeles, and the Jet Propulsion Laboratory/California Institute of Technology, funded by the National Aeronautics and Space Administration.

\hspace{1cm}{\it Facilities:} \facility{LBT/MODS1} \facility{MMT/Hectospec}

\appendix
\section{Appendix}
\begin{deluxetable}{lllc}
  \tablecaption{Foreground Dwarfs\label{tab:appendix}}
  \tabletypesize{\scriptsize}
  \tablecolumns{4}
  \tablewidth{0pt}
  \tablenum{A}
  \tablehead{\colhead{ID} & \colhead{Sp Type} & \colhead{Notes} & \colhead{Rank\tablenotemark{a}}}
  \startdata
  \bf{M31} & & & \\
  J003934.02+404714.2 & F0 &  & 2 \\
  J004107.40+405328.6 & A & HBA\tablenotemark{b} & 2 \\
  J004131.50+403917.8 & G0 &  & 1 \\
  J004144.76+402808.9 & G8 & composite? & 1 \\
  J004259.31+410629.1 & F2 & fgd in HII reg &  \\
  \tableline
  \bf{M33} & & & \\
  J013235.17+303331.6 & A8 & HBA\tablenotemark{b} & 1 \\
  J013320.57+304901.6 & F5 &  & 2 \\
  J013315.21+303727.0 & B+HII & hot star in HII reg &  \\
  J013331.15+304530.0 & A5 & He abs & 1 \\
  J013446.88+302620.9 & G &  & 2 \\
  \enddata
  \tablenotetext{a}{Ranks from \citet{drout_2009,drout_2012} specify if the source was a (1) ``highly likely'' or (2) ``possible'' supergiant.}
  \tablenotetext{b}{Indicates star is possibly a warm horizontal-branch A-type star.}
  \tablenotetext{\,}{(This table is available in its entirety in a machine-readable form in the online journal. A portion is shown here for guidance regarding its form and content.)}
\end{deluxetable}


\begin{thebibliography}{}
\expandafter\ifx\csname natexlab\endcsname\relax\def\natexlab#1{#1}\fi

\bibitem[{{Braun} {et~al.}(2009){Braun}, {Thilker}, {Walterbos}, \&
  {Corbelli}}]{braun_2009}
{Braun}, R., {Thilker}, D.~A., {Walterbos}, R.~A.~M., \& {Corbelli}, E. 2009,
  \apj, 695, 937

\bibitem[{{Cardelli} {et~al.}(1989){Cardelli}, {Clayton}, \&
  {Mathis}}]{cardelli_1989}
{Cardelli}, J.~A., {Clayton}, G.~C., \& {Mathis}, J.~S. 1989, \apj, 345, 245

\bibitem[{{de Jager}(1998)}]{dejager_1998AR}
{de Jager}, C. 1998, \aapr, 8, 145

\bibitem[{{de Jager} {et~al.}(1988){de Jager}, {Nieuwenhuijzen}, \& {van der
  Hucht}}]{dejager_1988}
{de Jager}, C., {Nieuwenhuijzen}, H., \& {van der Hucht}, K.~A. 1988, \aaps,
  72, 259

\bibitem[{{Draine} \& {Li}(2007)}]{draine_2007}
{Draine}, B.~T., \& {Li}, A. 2007, \apj, 657, 810

\bibitem[{{Drout} {et~al.}(2012){Drout}, {Massey}, \& {Meynet}}]{drout_2012}
{Drout}, M.~R., {Massey}, P., \& {Meynet}, G. 2012, \apj, 750, 97

\bibitem[{{Drout} {et~al.}(2009){Drout}, {Massey}, {Meynet}, {Tokarz}, \&
  {Caldwell}}]{drout_2009}
{Drout}, M.~R., {Massey}, P., {Meynet}, G., {Tokarz}, S., \& {Caldwell}, N.
  2009, \apj, 703, 441

\bibitem[{{Ekstr{\"o}m} {et~al.}(2012){Ekstr{\"o}m}, {Georgy}, {Eggenberger},
  {Meynet}, {Mowlavi}, {Wyttenbach}, {Granada}, {Decressin}, {Hirschi},
  {Frischknecht}, {Charbonnel}, \& {Maeder}}]{ekstrom_2012}
{Ekstr{\"o}m}, S., {Georgy}, C., {Eggenberger}, P., {et~al.} 2012, \aap, 537,
  A146

\bibitem[{{Fabricant} {et~al.}(2005){Fabricant}, {Fata}, {Roll}, {Hertz},
  {Caldwell}, {Gauron}, {Geary}, {McLeod}, {Szentgyorgyi}, {Zajac}, {Kurtz},
  {Barberis}, {Bergner}, {Brown}, {Conroy}, {Eng}, {Geller}, {Goddard},
  {Honsa}, {Mueller}, {Mink}, {Ordway}, {Tokarz}, {Woods}, {Wyatt}, {Epps}, \&
  {Dell'Antonio}}]{fabricant_2005}
{Fabricant}, D., {Fata}, R., {Roll}, J., {et~al.} 2005, \pasp, 117, 1411

\bibitem[{{Fabricant} {et~al.}(1998){Fabricant}, {Hertz}, {Szentgyorgyi},
  {Fata}, {Roll}, \& {Zajac}}]{fabricant_1998}
{Fabricant}, D.~G., {Hertz}, E.~N., {Szentgyorgyi}, A.~H., {et~al.} 1998, in
  Society of Photo-Optical Instrumentation Engineers (SPIE) Conference Series,
  Vol. 3355, Optical Astronomical Instrumentation, ed. S.~{D'Odorico}, 285--296

\bibitem[{{Flower}(1977)}]{flower_1977}
{Flower}, P.~J. 1977, \aap, 54, 31

\bibitem[{{Flower}(1996)}]{flower_1996}
---. 1996, \apj, 469, 355

\bibitem[{{Giannone}(1967)}]{giannone_1967}
{Giannone}, P. 1967, \zap, 65, 226

\bibitem[{{Gratier} {et~al.}(2010){Gratier}, {Braine}, {Rodriguez-Fernandez},
  {Schuster}, {Kramer}, {Xilouris}, {Tabatabaei}, {Henkel}, {Corbelli},
  {Israel}, {van der Werf}, {Calzetti}, {Garcia-Burillo}, {Sievers}, {Combes},
  {Wiklind}, {Brouillet}, {Herpin}, {Bontemps}, {Aalto}, {Koribalski}, {van der
  Tak}, {Wiedner}, {R{\"o}llig}, \& {Mookerjea}}]{gratier_2010}
{Gratier}, P., {Braine}, J., {Rodriguez-Fernandez}, N.~J., {et~al.} 2010, \aap,
  522, A3

\bibitem[{{Gustafsson} {et~al.}(2008){Gustafsson}, {Edvardsson}, {Eriksson},
  {J{\o}rgensen}, {Nordlund}, \& {Plez}}]{gustafsson_2008}
{Gustafsson}, B., {Edvardsson}, B., {Eriksson}, K., {et~al.} 2008, \aap, 486,
  951

\bibitem[{{Hartman} {et~al.}(2006){Hartman}, {Bersier}, {Stanek}, {Beaulieu},
  {Kaluzny}, {Marquette}, {Stetson}, \& {Schwarzenberg-Czerny}}]{hartman_2006}
{Hartman}, J.~D., {Bersier}, D., {Stanek}, K.~Z., {et~al.} 2006, \mnras, 371,
  1405

\bibitem[{{Hiltner} \& {Johnson}(1956)}]{hiltner_1956}
{Hiltner}, W.~A., \& {Johnson}, H.~L. 1956, \apj, 124, 367

\bibitem[{{Hora} {et~al.}(2004){Hora}, {Fazio}, {Allen}, {Ashby}, {Barmby},
  {Deutsch}, {Huang}, {Marengo}, {Megeath}, {Melnick}, {Pahre}, {Patten},
  {Smith}, {Wang}, {Willner}, {Hoffmann}, {Pipher}, {Forrest}, {McMurtry},
  {McCreight}, {McKelvey}, {McMurray}, {Moseley}, {Arendt}, {Mentzell}, {Marx},
  {Fixsen}, {Tollestrup}, {Eisenhardt}, {Stern}, {Gorjian}, {Bhattacharya},
  {Carey}, {Glaccum}, {Lacy}, {Lowrance}, {Laine}, {Nelson}, {Reach},
  {Stauffer}, {Surace}, {Wilson}, \& {Wright}}]{hora_2004}
{Hora}, J.~L., {Fazio}, G.~G., {Allen}, L.~E., {et~al.} 2004, in Society of
  Photo-Optical Instrumentation Engineers (SPIE) Conference Series, Vol. 5487,
  Optical, Infrared, and Millimeter Space Telescopes, ed. J.~C. {Mather},
  77--92

\bibitem[{{Humphreys} \& {Davidson}(1979)}]{humphreys_1979}
{Humphreys}, R.~M., \& {Davidson}, K. 1979, \apj, 232, 409

\bibitem[{{Humphreys} \& {Davidson}(1994)}]{humphreys_1994}
---. 1994, \pasp, 106, 1025

\bibitem[{{Humphreys} {et~al.}(2013){Humphreys}, {Davidson}, {Grammer},
  {Kneeland}, {Martin}, {Weis}, \& {Burggraf}}]{humphreys_2013}
{Humphreys}, R.~M., {Davidson}, K., {Grammer}, S., {et~al.} 2013, \apj, 773, 46

\bibitem[{{Humphreys} {et~al.}(2002){Humphreys}, {Davidson}, \&
  {Smith}}]{humphreys_2002}
{Humphreys}, R.~M., {Davidson}, K., \& {Smith}, N. 2002, \aj, 124, 1026

\bibitem[{{Humphreys} {et~al.}(2015){Humphreys}, {Martin}, \&
  {Gordon}}]{humphreys_2015}
{Humphreys}, R.~M., {Martin}, J.~C., \& {Gordon}, M.~S. 2015, \pasp, 127, 347

\bibitem[{{Humphreys} {et~al.}(2014){Humphreys}, {Weis}, {Davidson}, {Bomans},
  \& {Burggraf}}]{humphreys_2014}
{Humphreys}, R.~M., {Weis}, K., {Davidson}, K., {Bomans}, D.~J., \& {Burggraf},
  B. 2014, \apj, 790, 48

\bibitem[{{Humphreys} {et~al.}(1997){Humphreys}, {Smith}, {Davidson}, {Jones},
  {Gehrz}, {Mason}, {Hayward}, {Houck}, \& {Krautter}}]{humphreys_1997}
{Humphreys}, R.~M., {Smith}, N., {Davidson}, K., {et~al.} 1997, \aj, 114, 2778

\bibitem[{{Humphreys} {et~al.}(2006){Humphreys}, {Jones}, {Polomski},
  {Koppelman}, {Helton}, {McQuinn}, {Gehrz}, {Woodward}, {Wagner}, {Gordon},
  {Hinz}, \& {Willner}}]{humphreys_2006}
{Humphreys}, R.~M., {Jones}, T.~J., {Polomski}, E., {et~al.} 2006, \aj, 131,
  2105

\bibitem[{{Johnson}(1958)}]{johnson_1958}
{Johnson}, H.~L. 1958, Lowell Observatory Bulletin, 4, 37

\bibitem[{{Jones} {et~al.}(1993){Jones}, {Humphreys}, {Gehrz}, {Lawrence},
  {Zickgraf}, {Moseley}, {Casey}, {Glaccum}, {Koch}, {Pina}, {Jones}, {Venn},
  {Stahl}, \& {Starrfield}}]{jones_1993}
{Jones}, T.~J., {Humphreys}, R.~M., {Gehrz}, R.~D., {et~al.} 1993, \apj, 411,
  323

\bibitem[{{Kaluzny} {et~al.}(1998){Kaluzny}, {Stanek}, {Krockenberger},
  {Sasselov}, {Tonry}, \& {Mateo}}]{kaluzny_1998}
{Kaluzny}, J., {Stanek}, K.~Z., {Krockenberger}, M., {et~al.} 1998, \aj, 115,
  1016

\bibitem[{{King} {et~al.}(1998){King}, {Walterbos}, \& {Braun}}]{king_1998}
{King}, N.~L., {Walterbos}, R.~A.~M., \& {Braun}, R. 1998, \apj, 507, 210

\bibitem[{{Knapp} {et~al.}(1973){Knapp}, {Rose}, \& {Kerr}}]{knapp_1973}
{Knapp}, G.~R., {Rose}, W.~K., \& {Kerr}, F.~J. 1973, \apj, 186, 831

\bibitem[{{Levesque} {et~al.}(2006){Levesque}, {Massey}, {Olsen}, {Plez},
  {Meynet}, \& {Maeder}}]{levesque_2006}
{Levesque}, E.~M., {Massey}, P., {Olsen}, K.~A.~G., {et~al.} 2006, \apj, 645,
  1102

\bibitem[{{Macri} {et~al.}(2001){Macri}, {Stanek}, {Sasselov}, {Krockenberger},
  \& {Kaluzny}}]{macri_2001}
{Macri}, L.~M., {Stanek}, K.~Z., {Sasselov}, D.~D., {Krockenberger}, M., \&
  {Kaluzny}, J. 2001, \aj, 121, 870

\bibitem[{{Massey} {et~al.}(2007{\natexlab{a}}){Massey}, {McNeill}, {Olsen},
  {Hodge}, {Blaha}, {Jacoby}, {Smith}, \& {Strong}}]{massey_2007c}
{Massey}, P., {McNeill}, R.~T., {Olsen}, K.~A.~G., {et~al.} 2007{\natexlab{a}},
  \aj, 134, 2474

\bibitem[{{Massey} {et~al.}(2016){Massey}, {Neugent}, \& {Smart}}]{massey_2016}
{Massey}, P., {Neugent}, K.~F., \& {Smart}, B.~M. 2016, ArXiv e-prints,
  arXiv:1604.00112

\bibitem[{{Massey} {et~al.}(2007{\natexlab{b}}){Massey}, {Olsen}, {Hodge},
  {Jacoby}, {McNeill}, {Smith}, \& {Strong}}]{massey_2007b}
{Massey}, P., {Olsen}, K.~A.~G., {Hodge}, P.~W., {et~al.} 2007{\natexlab{b}},
  \aj, 133, 2393

\bibitem[{{Massey} {et~al.}(2006){Massey}, {Olsen}, {Hodge}, {Strong},
  {Jacoby}, {Schlingman}, \& {Smith}}]{massey_2006}
---. 2006, \aj, 131, 2478

\bibitem[{{Massey} {et~al.}(2009){Massey}, {Silva}, {Levesque}, {Plez},
  {Olsen}, {Clayton}, {Meynet}, \& {Maeder}}]{massey_2009}
{Massey}, P., {Silva}, D.~R., {Levesque}, E.~M., {et~al.} 2009, \apj, 703, 420

\bibitem[{{Mauron} \& {Josselin}(2011)}]{mauron_2011}
{Mauron}, N., \& {Josselin}, E. 2011, \aap, 526, A156

\bibitem[{{McQuinn} {et~al.}(2007){McQuinn}, {Woodward}, {Willner}, {Polomski},
  {Gehrz}, {Humphreys}, {van Loon}, {Ashby}, {Eicher}, \&
  {Fazio}}]{mcquinn_2007}
{McQuinn}, K.~B.~W., {Woodward}, C.~E., {Willner}, S.~P., {et~al.} 2007, \apj,
  664, 850

\bibitem[{{Mould} {et~al.}(2008){Mould}, {Barmby}, {Gordon}, {Willner},
  {Ashby}, {Gehrz}, {Humphreys}, \& {Woodward}}]{mould_2008}
{Mould}, J., {Barmby}, P., {Gordon}, K., {et~al.} 2008, \apj, 687, 230

\bibitem[{{Nieuwenhuijzen} {et~al.}(2012){Nieuwenhuijzen}, {De Jager}, {Kolka},
  {Israelian}, {Lobel}, {Zsoldos}, {Maeder}, \& {Meynet}}]{nieu_2012}
{Nieuwenhuijzen}, H., {De Jager}, C., {Kolka}, I., {et~al.} 2012, \aap, 546,
  A105

\bibitem[{{Oudmaijer}(1998)}]{oud_1998}
{Oudmaijer}, R.~D. 1998, \aaps, 129, 541

\bibitem[{{Oudmaijer} {et~al.}(1996){Oudmaijer}, {Groenewegen}, {Matthews},
  {Blommaert}, \& {Sahu}}]{oud_1996}
{Oudmaijer}, R.~D., {Groenewegen}, M.~A.~T., {Matthews}, H.~E., {Blommaert},
  J.~A.~D.~L., \& {Sahu}, K.~C. 1996, \mnras, 280, 1062

\bibitem[{{Rachford} {et~al.}(2009){Rachford}, {Snow}, {Destree}, {Ross},
  {Ferlet}, {Friedman}, {Gry}, {Jenkins}, {Morton}, {Savage}, {Shull},
  {Sonnentrucker}, {Tumlinson}, {Vidal-Madjar}, {Welty}, \&
  {York}}]{rachford_2009}
{Rachford}, B.~L., {Snow}, T.~P., {Destree}, J.~D., {et~al.} 2009, \apjs, 180,
  125

\bibitem[{{Savage} \& {Jenkins}(1972)}]{savage_1972}
{Savage}, B.~D., \& {Jenkins}, E.~B. 1972, \apj, 172, 491

\bibitem[{{Shenoy} {et~al.}(2016){Shenoy}, {Humphreys}, {Jones}, {Marengo},
  {Gehrz}, {Helton}, {Hoffmann}, {Skemer}, \& {Hinz}}]{shenoy_2016}
{Shenoy}, D., {Humphreys}, R.~M., {Jones}, T.~J., {et~al.} 2016, \aj, 151, 51

\bibitem[{{Sholukhova} {et~al.}(2015){Sholukhova}, {Bizyaev}, {Fabrika},
  {Sarkisyan}, {Malanushenko}, \& {Valeev}}]{sholukhova_2015}
{Sholukhova}, O., {Bizyaev}, D., {Fabrika}, S., {et~al.} 2015, \mnras, 447,
  2459

\bibitem[{{Skrutskie} {et~al.}(2006){Skrutskie}, {Cutri}, {Stiening},
  {Weinberg}, {Schneider}, {Carpenter}, {Beichman}, {Capps}, {Chester},
  {Elias}, {Huchra}, {Liebert}, {Lonsdale}, {Monet}, {Price}, {Seitzer},
  {Jarrett}, {Kirkpatrick}, {Gizis}, {Howard}, {Evans}, {Fowler}, {Fullmer},
  {Hurt}, {Light}, {Kopan}, {Marsh}, {McCallon}, {Tam}, {Van Dyk}, \&
  {Wheelock}}]{skrutskie_2006}
{Skrutskie}, M.~F., {Cutri}, R.~M., {Stiening}, R., {et~al.} 2006, \aj, 131,
  1163

\bibitem[{{Smartt}(2015)}]{smartt_2015}
{Smartt}, S.~J. 2015, \pasa, 32, 16

\bibitem[{{Smartt} {et~al.}(2009){Smartt}, {Eldridge}, {Crockett}, \&
  {Maund}}]{smartt_2009}
{Smartt}, S.~J., {Eldridge}, J.~J., {Crockett}, R.~M., \& {Maund}, J.~R. 2009,
  \mnras, 395, 1409

\bibitem[{{Suh}(1999)}]{suh_1999}
{Suh}, K.-W. 1999, \mnras, 304, 389

\bibitem[{{Thompson} {et~al.}(2009){Thompson}, {Prieto}, {Stanek}, {Kistler},
  {Beacom}, \& {Kochanek}}]{thompson_2009}
{Thompson}, T.~A., {Prieto}, J.~L., {Stanek}, K.~Z., {et~al.} 2009, \apj, 705,
  1364

\bibitem[{{Treffers} \& {Cohen}(1974)}]{treffers_1974}
{Treffers}, R., \& {Cohen}, M. 1974, \apj, 188, 545

\bibitem[{{Valeev} {et~al.}(2010){Valeev}, {Sholukhova}, \&
  {Fabrika}}]{valeev_2010}
{Valeev}, A.~F., {Sholukhova}, O.~N., \& {Fabrika}, S.~N. 2010, Astrophysical
  Bulletin, 65, 140

\bibitem[{{Vink}(2012)}]{vink_2012}
{Vink}, J.~S. 2012, in Astrophysics and Space Science Library, Vol. 384, Eta
  Carinae and the Supernova Impostors, ed. K.~{Davidson} \& R.~M. {Humphreys},
  221

\bibitem[{{Woolf} \& {Ney}(1969)}]{woolf_1969}
{Woolf}, N.~J., \& {Ney}, E.~P. 1969, \apjl, 155, L181

\bibitem[{{Wright} {et~al.}(2010){Wright}, {Eisenhardt}, {Mainzer}, {Ressler},
  {Cutri}, {Jarrett}, {Kirkpatrick}, {Padgett}, {McMillan}, {Skrutskie},
  {Stanford}, {Cohen}, {Walker}, {Mather}, {Leisawitz}, {Gautier}, {McLean},
  {Benford}, {Lonsdale}, {Blain}, {Mendez}, {Irace}, {Duval}, {Liu}, {Royer},
  {Heinrichsen}, {Howard}, {Shannon}, {Kendall}, {Walsh}, {Larsen}, {Cardon},
  {Schick}, {Schwalm}, {Abid}, {Fabinsky}, {Naes}, \& {Tsai}}]{wright_2010}
{Wright}, E.~L., {Eisenhardt}, P.~R.~M., {Mainzer}, A.~K., {et~al.} 2010, \aj,
  140, 1868

\end{thebibliography}
\end{document}